\documentclass[aip,
amsmath,amssymb,floatfix,
reprint
]{revtex4-2}
\usepackage{lipsum,amsmath,amssymb,CJK}
\usepackage{url}
\usepackage{multirow}
\usepackage{graphicx}% Include figure files
\usepackage{dcolumn}% Align table columns on decimal point
\usepackage{bm}% bold math
\usepackage[utf8]{inputenc}
\usepackage[T1]{fontenc}
\usepackage{mathptmx}

\begin{document}
\preprint{AIP/123-QED}
\begin{CJK*}{UTF8}{}
\CJKfamily{mj}
\title[Power-grid stability predictions using transferable machine learning]
{Power-grid stability predictions using transferable machine learning}

\author{Seong-Gyu Yang (양성규)}
\affiliation{Asia Pacific Center for Theoretical Physics, Pohang 37673, Republic of Korea}
\affiliation{Department of Physics, Sungkyunkwan University, Suwon 16419, Republic of Korea}
\author{Beom Jun Kim (김범준)}
\affiliation{Department of Physics, Sungkyunkwan University, Suwon 16419, Republic of Korea}
\author{Seung-Woo Son (손승우)}
\email{sonswoo@hanyang.ac.kr}
\affiliation{Asia Pacific Center for Theoretical Physics, Pohang 37673, Republic of Korea}
\affiliation{Department of Applied Physics, Center for Bionano Intelligence Education and Research, Hanyang University, Ansan 15588, Republic of Korea}
\author{Heetae Kim (김희태)}
\email{hkim@kentech.ac.kr}
\affiliation{Department of Energy Technology, Korea Institute of Energy Technology, Naju 58330, Republic of Korea}
\affiliation{Instituto de Data Science, Universidad del Desarrollo, Santiago 7610658, Chile}
\date{\today}

\begin{abstract}
Complex network analyses have provided clues to improve power-grid stability with the help of numerical models.
The high computational cost of numerical simulations, however, has inhibited the approach, especially when it deals with the dynamic properties of power grids such as frequency synchronization.
In this study, we investigate machine learning techniques to estimate the stability of power-grid synchronization.
We test three different machine learning algorithms---random forest, support vector machine, and artificial neural network---training them with two different types of synthetic power grids consisting of homogeneous and heterogeneous input-power distribution, respectively.
We find that the three machine learning models better predict the synchronization stability of power-grid nodes when they are trained with the heterogeneous input-power distribution than the homogeneous one.
With the real-world power grids of Great Britain, Spain, France, and Germany, we also demonstrate that the machine learning algorithms trained on synthetic power grids are transferable to the stability prediction of the real-world power grids, which implies the prospective applicability of machine learning techniques on power-grid studies.
\end{abstract}

\maketitle
\end{CJK*}

\begin{quotation}
The increasing capacity of renewable power generation brings about frequent fluctuations in a power grid's operational frequency, which consequently threatens the power-grid stability.
Thus, it is becoming more important to keep the power system controlled as well as to predict its stability.
In this paper, we aim at the prediction of the synchronization recovery of power-grid nodes using machine learning techniques.
As large data sets to train machine learning models, we first generate synthetic power grids~\cite{netgen}.
We find that three machine learning models---random forest (RF), support vector machine (SVM), and artificial neural network (NN)---well predict the stability of the synthetic power-grid nodes.
Furthermore, we successfully apply the machine learning algorithms to the stability analysis of real power grids~\cite{EU,German1}---British, French, Spanish, and German power grids---and find that our models are transferable to different power-grid structures with various system sizes.
\end{quotation}

\section{INTRODUCTION}
\label{sec:intro}
The increasing of renewable capacities can destabilize a power grid due to the high dependency of renewable energy sources on external dynamic factors such as weather conditions~\cite{PV_unstable,wind_unstable}.
Previous studies have focused on the topological properties of power grids~\cite{complex_network1,complex_network2,topological,topological_Korea} and their structural vulnerability~\cite{topological,topological_Korea,vulnerability1,vulnerability2,vulnerability3,vulnerability4,vulnerability5,vulnerability_topology}.
However, it is also required to estimate the stability of power grids in terms of the phase and frequency dynamics of power grids.
The dynamics of an alternating current (AC) power system can be approximated by the second-order Kuramoto model~\cite{swing1,Sync_review,Sync_review2} assuming the constant voltage amplitude and neglecting the energy loss in power transmission.
Accordingly, the stability evaluation of power systems considering the dynamics of AC power has been studied rigorously in recent works~\cite{not_only_topology,Structural_dynamical,BS2,BS3,cascading,EU,GB_decentral,EU,Turkish,German2,vulnerability_AC,first_pert_time}.

The dynamic model of an AC power system enables us to estimate the stability of the system through numerical simulations.
The synchronization stability of a power-grid node is measured by the basin stability, which is defined as the relative volume of the basin of attraction in the state space when perturbing the node~\cite{BS_introduced1,BS_introduced2}.
The precise estimation of the basin of attraction is difficult because the area often exhibits a fractal-like pattern~\cite{BS3}.
Thus, the Monte Carlo method is one of the options to numerically measure the basin stability, for which we randomly perturb each node of a power system.

However, the high computational cost in the numerical simulations makes the stability evaluation of power systems difficult.
It is because the stability depends not only on the network topology~\cite{stability1,impact_network,DeadEnds,k_core,CoI} but also on the parameters related to the dynamics of AC power~\cite{GB_decentral,BS2,BS3,damping_Motter} that require multiple combinations of various test sets.
Thus, the estimation of the power-grid stability exploring the parameter space is a challenging problem, for which we have tried to apply machine learning (ML) techniques.
In addition, from a practical point of view, the local failure of the power system can cause a nationwide, sometimes continental scale disaster in a couple of minutes~\cite{Fail_Time}.
To prevent the disaster, instant and appropriate remedial actions should be taken in a few seconds or minutes.
The detection of the local failure and the result prediction via numerical simulations take lots of time and cannot nip the large scale disaster in the bud.

Recently, ML approaches have been used in various fields of physics to reduce the computational costs.
For example, the Monte Carlo simulations of glass systems~\cite{ML_MC} and molecular dynamics simulation~\cite{ML_MD} have been boosted with a ML technique.
ML is also applied in the prediction of nonlinear dynamic systems~\cite{ML_RC,ML_ComplexSystem,ML_Forecast,motivation}.
Rodrigues~\emph{et al.} have shown that the applicability of ML techniques to predict the size of epidemic outbreaks and the final state of the coupled oscillator system from the given network topologies and initial conditions~\cite{motivation}.

In the previous studies, ML techniques have been utilized to predict the power outages, which are induced by natural disasters~\cite{ML_Hurricane, ML_Failure} but the effect of power-grid topology is not considered in these studies.
The prediction of the power failure is treated as the binary classification problem in those studies.
The logistic regression method is used to predict the system failure induced by a hurricane on the artificial data set using the distance from the hurricane center and the wind speed as input features~\cite{ML_Hurricane}.
The artificial neural network method is also used to predict the power outage in the North American power systems considering the drought and hurricane effects and also the misoperation rate of a power control system~\cite{ML_Failure}.
The ML techniques are also applied to power system stability assessment~\cite{ML_PG_stability,ML_PQ_review2}.
The performance of artificial neural networks, support vector machines, decision trees, and other ML techniques is compared in the review paper.
However, most of the research studies in the review paper studied the IEEE test system but not the real power-grid topology.

Furthermore, many researchers have utilized ML techniques to estimate the power quality and classify the power quality disturbances~\cite{ML_PQ_review1,ML_PQ_review2,ML_PQ_review3,ML_PQ_review4}.
The standards on power quality, such as IEC 61000-4-30 standard, IEEE 1159 standard, and EN 50160 standard, are developed and issued by international organizations.
Using ML models to classify the power quality disturbances based on the standards, features of the disturbances are extracted from the original signal and processed signal.
In this stage, several signal processing techniques, including Fourier transform and wavelet transform, are used.
The extracted features are fed to ML classifiers, such as artificial neural networks, support vector machines, decision trees, fuzzy logic, etc., to recognize the type of disturbances and the performance of many different ML techniques are compared in review papers~\cite{ML_PQ_review1,ML_PQ_review2,ML_PQ_review3,ML_PQ_review4}.
In these studies, however, only the voltage and current signal data are considered ignoring the topological effect on power systems.

Although ML is one of the effective methods achieving breakthroughs in many problems, it has some drawbacks when applied to the stability prediction of power grids.
One drawback is the large-size data set that is essential to properly train ML models.
Although there are several open data sets of real-world power grids~\cite{Chile_data,PanTaGruEl,German1,ENTSO-E,ENTSO-E_review}, the size and type of power-grid data are still limited.
Moreover, when ML models are trained on a specific power-grid topology, they become too optimized (over-fitted) to the topology so that they are hardly applicable to the other power-grid topologies.
These limitations, deficient open data sets and high dependency due to the training on a single power-grid topology, are a challenge to develop transferable ML models which can predict the power-grid stability on any given topology.

In the present paper, we investigate the applicability and transferability of ML algorithms to predict the power-grid stability.
The power grid recovery from random perturbation can be regarded as the binary classification problem, and we investigate three different ML algorithms---random forest (RF), support vector machine (SVM), and artificial neural network (NN)---to predict whether a power grid recovers its synchronous state against random perturbations or not.
First, we construct the training data sets numerically simulating the synthetic power grids that are generated from the random growth model of power grids~\cite{netgen}.
Two different power distribution scenarios---\emph{homogeneous} (HOM) and \emph{heterogeneous} (HET)---are also introduced to estimate the performance of ML algorithms for the different levels of complexity.
After constructing the training data sets, in the ML training step, we choose six topological features and three dynamical features to train the transferable ML models~\cite{motivation}.
After we train the ML models on the synthetic power grids with the power distributions, we apply the ML models to predict the stability of the real-world power grids---British, French, Spanish, and German---which have different network topologies with different system sizes.
Different from the previous studies~\cite{ML_PG_stability,ML_PQ_review2}, we train three ML models on synthetic power grids and validate the prediction performance on the real power grids to test the transferability of the ML models.

This paper is organized as follows:
in Sec.~\ref{ssec:model}, we present the dynamic model of AC power systems which is the second-order Kuramoto type.
The details of the three ML algorithms (Sec.~\ref{ssec:ML_algorithms}) and the description of how we construct the training data set using the random growth model (Sec.~\ref{ssec:data_synthesis}) follow.
It is described in Sec.~\ref{ssec:input_features}, in which topological measures are used as input features, and the description of the test data sets for real power grids is given in Sec.~\ref{ssec:real_data}.
We report the prediction performance of our ML models for the training data set and for the test data sets in Sec.~\ref{sec:results} followed by the conclusion and discussions of this paper.

%%%%%%%%%%%%%%%%%%%%%%%%%%%%%%%%%%%
\section{METHODS}
\label{sec:methods}
\subsection{Dynamic model of AC power systems: the second-order Kuramoto model}
\label{ssec:model}
%%% Swing equation
The synchronous phase dynamics of the AC power-grid system is usually described by the following second-order Kuramoto model, which is derived from the energy conservation law in power-grid systems~\cite{swing1,swing2,Sync_review,Sync_review2},
\begin{eqnarray}
\dot{\theta}_i &=& \omega_i, \nonumber \\
\dot{\omega}_i &=& P_i - \alpha_i \omega_i + \sum_{j}K_{ij} \sin(\theta_j - \theta_i),
\label{eq:swing}
\end{eqnarray}
where $\theta_i$ is the phase angle variable and $\omega_i$ is the angular frequency of node $i$ in the reference frame of the rated frequency.
The energy dissipation term is written as a multiplication of $\omega_i$ and the damping coefficient $\alpha_i > 0$.
$K_{ij}$ denotes the coupling strength between nodes $i$ and $j$, which corresponds to the electric capacity of a transmission line between nodes $i$ and $j$.
$P_i$ is the amount of net power production in node $i$, which is positive (negative) if node $i$ is a producer (consumer).
For energy conservation of the whole system, the total net power production over the networks should be zero, i.e., $\sum_i P_i =0$.
Different from the conventional Kuramoto model, the angular frequency $\omega_i$ plays as an indicator of synchronization to the rated frequency of the power-grid system.
In the power grid, a phase-locked state is desired. In that state, all the oscillators move together with the same rated frequency $50$ or $60$ Hz but can have different voltaic phases.
The zero angular frequency for all nodes denotes that the whole system reaches the synchronization of the rated AC frequency.

%% ML algorithms
\subsection{Three ML algorithms}
\label{ssec:ML_algorithms}
We design the stability prediction from random perturbations as the binary classification problem.
Then, we use three ML algorithms to predict the stability of synthetic power grids.
One is RF, which uses the ensemble of decision trees~\cite{RF,RF2}; the second is SVM, which is the method to find a hyperplane separating the data points into differently labeled groups in data space~\cite{SVM,SVM2}; and the last is NN, which is introduced by mimicking the neuron connection~\cite{ML_book,NN_dropout,ML_book_Goodfellow}.
Those ML algorithms are three of the frequently used basic classification algorithms~\cite{ML_book,ML_book_Goodfellow}, and we use all three in this paper.

For RF, we use $500$ different decision trees, and each decision tree is trained with a different bootstrap data sample~\cite{RF,ML_book}.
We use the Gaussian radial basis function kernel for SVM~\cite{SVM2,ML_book} which is the method to make the hyperplane nonlinear in the input feature space.
The NN model consists of an input layer with a bias node, four hidden layers of $10\times20\times 20\times 10$ structure with an additional bias node for each hidden layer, and a dropout rate $0.1$, and an output layer~\cite{NN_dropout,ML_book,ML_book_Goodfellow}.
Rectified linear unit (ReLU) activation function in the hidden layers and the softmax activation function in the output layer are applied.
We use the binary cross entropy as the objective function~\cite{ML_book,ML_book_Goodfellow} and Adam optimizer~\cite{Adam} to find the minimum value of the objective function.
We utilize the python scikit-learn package~\cite{ML_book,scikit-learn,scikit-learn_url} to construct RF and SVM models and the pytorch package~\cite{pytorch,pytorch_url} to build NN models.

%%%%%%%%%%%%%%%%%%%%%%%%%%%%%%%%%%%
%% Power grid generation
\subsection{Training on synthetic power grids}
\label{ssec:data_synthesis}
We generate synthetic power grids consisting of $N=500$ nodes using a random growth model in Ref.~\onlinecite{netgen} with $(p,q,r,s)=(0.03,0.44,0.3,0.28)$, where $p$ is the link wiring probability, $q$ is the probability for wiring redundancy line, $r$ is the parameter related with the redundancy/cost trade-off, and $s$ is the probability for splitting the existing link.
We choose the same parameters, which are used to generate power grids similar to the Western US power grid in topological properties.
In each synthetic power grid, we assign two different power distributions.
(i) HOM distribution: randomly chosen $N/2(=250)$ nodes are assigned as producers ($P_i =+1$) and the other half as consumers ($P_i =-1$).
The absolute value of net power production is homogeneous as $| P_i | =1$ for all nodes.
(ii) HET distribution: randomly chosen $N/10(=50)$ nodes are the large producers (consumers) with net power production $P_i=2 (-2)$ and other $2N/5(=200)$ nodes are the small producers (consumers) having $P_i=0.2 (-0.2)$.
In the HET distribution, we consider a few large producers/consumers and many small producers/consumers, which is more similar to real power grids than the HOM distribution.

%% Numerical integration
On the generated synthetic power-grid structure, we numerically integrate Eq.~\eqref{eq:swing} using the fourth-order Runge-Kutta method with the integration time step size $\Delta t=10^{-3}$ for $5\times 10^5$ time steps.
The coupling strength for every connected pair of nodes $i$ and $j$ is chosen as $K_{ij}=K=10$, which is large enough for the system reaching the synchronous state as the stable fixed point.
The damping parameter is $\alpha_i=\alpha=0.1$ for all nodes.

After the power grid reaches a synchronous state, we perturb the state of node $i$ changing its phase and angular frequency $(\theta_i,\omega_i)$ to the new state $(\omega_\text{pert}, \theta_\text{pert})$, which are randomly chosen from the uniform distribution in the range of $[-50,50]\times [-\pi,\pi ]$.
We apply ten random perturbations on each node to build the data sets for the synthetic power grids and then put the label for each trial as ``$+1$'' if the system recovers its synchronous state from the given perturbation or ``$-1$'' for the opposite cases.

The whole data set consists of nine input features and the label of ``$+1$'' or ``$-1$,'' which denotes whether the system recovers from the given perturbation or not.
We remove the label bias in the data sets obtained from the numerical simulation on the synthetic power grids balancing the number of data points labeled with ``$+1$'' and ``$-1$,'' and divide the data sets into training ($80\%$) and test data set ($20\%$) for each power distributions.
We train the ML models with training data sets and estimate the prediction performance with the test data sets.
All performance measures are averaged over ten differently sampled data sets from the whole data sets.
The performance measures of the ML models on synthetic power grids are presented in Sec.~\ref{ssec:synthetic_results}.

%%%%%%%%%%%%%%%%%%%%%%%%%%%%%%%%%%%
%% Input features
\subsection{Input features}
\label{ssec:input_features}
Power-grid stability depends not only on the AC dynamics of voltage but also on the network topology of power producers and consumers~\cite{impact_network,Structural_dynamical,BS2,BS3,CoI}.
Thus, we choose three dynamic features---net power production ($P$), phase perturbation ($\theta_\text{pert}$), and frequency perturbation ($\omega_\text{pert}$)---with six topological features---degree ($k$), $k$-core ($\text{core}$), eigenvector centrality ($\text{EC}$), closeness centrality ($\text{CC}$), betweenness centrality ($\text{BC}$)~\cite{network_newman,network_barabasi}, and companionship inconsistency ($\text{CoI}$)~\cite{CoI,CoI_network}---as input features of the ML model training.

The degree of a node $k$ is the number of its neighbor.
The $k$-core denotes the largest subnetwork where each node has degree $k$ at least and a node has the core value of $k$ if the node belongs to the $k$-core but does not to $(k+1)$-core.
In the previous study, Yang \emph{et al.} showed that the core assigned in a transmission line has a high correlation with the vulnerability of that line in the US--South Canada power grid~\cite{k_core}.
The EC measures the importance of each node related to the eigenvector $\vec{\bf x}$ corresponding to the largest eigenvalue $\lambda$ of the network adjacency matrix $\mathbf{A}$, i.e.,
\begin{eqnarray}
\mathbf{A}\vec{\bf x} = \lambda \vec{\bf x}.
\label{eq:EC}
\end{eqnarray}
The $i$th component of $\vec{\bf x}$ gives the EC of node $i$.
The CC of node $i$ is the reciprocal of the average shortest path length from node $i$ to other nodes such that 
\begin{eqnarray}
\text{CC}_i = \frac{N-1}{\sum_{j\neq i}d_{ij}},
\label{eq:CC}
\end{eqnarray}
where $N$ is the system size and $d_{ij}$ is the network distance between nodes $i$ and $j$.
The BC of node $i$ measures how many times the node $i$ lies on the shortest path between all pairs of other nodes in the network,
\begin{eqnarray}
\text{BC}_i=\sum_{j,l} \frac{s'_{jl}(i)}{s_{jl}},
\label{eq:BC}
\end{eqnarray}
where $s_{jl}$ is the number of shortest paths between nodes $j$ and $l$ and $s'_{jl}(i)$ is the number of the shortest paths between nodes $j$ and $l$, which pass through node $i$.
The nodes connecting two different groups or clusters usually have high BC values.
The CoI of node $i$ measures the level of inconsistency while node $i$ belongs to a community~\cite{CoI_network,CoI}.
In the Chilean power grid, nodes having low CoI value tend to have a wide basin stability transition window, which can be utilized as an indicator of system failure~\cite{CoI}. 
Degree $k$ and the dynamical features have local information of each node but the other topological measures reflect the global information of power-grid topology.

%%%%%%%%%%%%%%%%%%%%%%%%%%%%%%%%%%%
%% Power grid generation
\subsection{Real power-grid data}
\label{ssec:real_data}

\begin{figure*}[t]
\includegraphics[width=0.80\linewidth]{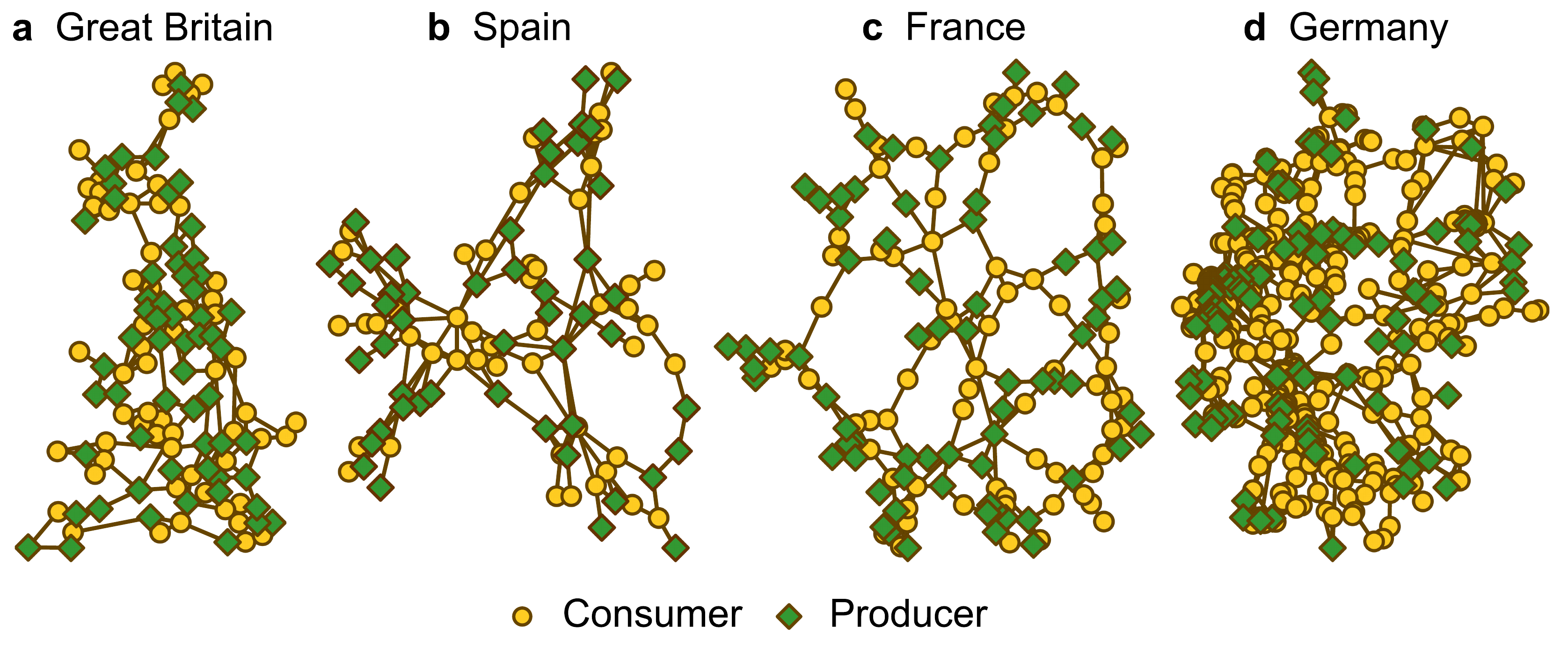}
\centering
\caption{
Network topologies of four European power grids.
The yellow circles (green rhombi) denote the consumers (producers).
The network sizes of real power grids are $N_\text{GB}=120$, $N_\text{ES}=98$, $N_\text{FR}=146$, and $N_\text{DE}=438$.
(a)-(c) For British, French, and Spanish power grids, we generate the power grids with the power distributions used in the synthetic power-grid generation due to the absence of the power generation and consumption data.
(d) For the German power grid, we utilize the real power generation and consumption data that are included in the open data set of ELMOD-DE.
In the German power grid, $113$ nodes are producers and $325$ nodes are consumers.
}
\label{fig:EU_PG}
\end{figure*}

To test the transferability of our ML models, we prepare the data sets for different power-grid topologies in the real world.
We utilize the open public data set of British, Spanish, French, and German power grids for the network structures~\cite{EU,German1}.
Figure~\ref{fig:EU_PG} displays the four European power grids, where the yellow circles denote the consumers and green rhombi denote the producers.
The system sizes of the real power grids are different and their structures show different features. 

The power generation and consumption information of each node is not available for British, Spanish, and French power grids.
Thus, we randomly assign the net power production $P_i$ for each node with two power distributions, which are used in the synthetic power grids (details in Sec.~\ref{ssec:data_synthesis}).
The system sizes of three European power grids are $N_\text{GB}=120$ (Great Britain), $N_\text{FR}=146$ (France), and $N_\text{ES}=98$ (Spain).
For three European power grids, we assign the same value as in Ref.~\onlinecite{EU} to the transmission capacity $K$. 
With the damping parameter $\alpha=0.1$, we numerically integrate Eq.~\eqref{eq:swing} on each power grid.
If the system does not reach the synchronous state, we newly distribute the net power production $P$ for every node and integrate Eq.~\eqref{eq:swing} again until the system becomes stable. 

For the German power grid, the net power production data of each node are available in the ELMOD-DE data set~\cite{German1}.
We utilize the net power production data to simulate the German power grid.
The system size of the German power grid is $N_\text{DE}=438$.
The net power production of node $i$ is determined as
\begin{eqnarray}
P_i = \gamma~ C_i - D_i + I_i - E_i,
\label{eq:P_German}
\end{eqnarray}
where $C_i$ is the maximum power generation capacity of power plant node $i$, $D_i$ is the power demand of node $i$, and $\gamma$ is the operation rate of power plants, which is fixed as $\gamma=0.4$ in our simulation.
$I_i (E_i)$ is the annual average of power import (export) of node $i$ which has cross-border transmission line.
In the ELMOD-DE raw data~\cite{German1}, there are hourly power import and export data for a year in the unit of electric energy ($\text{GWh}$) and we change them in the unit of electric power ($\text{GW}$) averaging the annual data and dividing them by $3600$ s.
For the power balance, i.e., $\sum_i P_i =0$, we distribute the net power demand equally to other nodes where the net power production is zero.
Differently from Ref.~\onlinecite{German2}, we consider the power generation not only from fossil fuels but also from the renewable energy sources including run-of-river, biomass, geothermal, photovoltaic, onshore, and offshore wind power.
We also consider the annual average power import and export near country borders.
We find the minimum $K$ in which the power grid operates stably and fix all electric capacity of transmission lines as $K=0.49$ ($\approx$ 0.62 GW).
Even for the minimum $K$ we find, the average basin stability of the German power grid shows a high value of $0.869$.
We adequately rescale all variables to dimensionless quantities~\cite{swing1,swing2} using the damping parameter $\alpha=0.1$ and the angular momentum of rotors rotating in a rated frequency of $50$ Hz.
We assume that each node's moment of inertia is $40 \times 10^3~\text{kg}\,\text{m}^2$, which corresponds to the inertia of rotors in $400$ MW power plants~\cite{German2,DeadEnds}.

After the system reaches the synchronous steady state, we build the test data sets for the real power grids applying $50$ random perturbations to each node on the real power-grid topologies and collect the results of recovery from the perturbations.
The real power grids deliver the electric power stably as designed; thus, the power grids show high basin stability.
Thus, we do not make the number of positively and negatively labeled data equal.
All data points are used to estimate the performance of our ML models on the real power grids.
The results on the real power grids are presented in Sec.~\ref{ssec:real_results}.

%%%%%%%%%%%%%%%%%%%%%%%%%%%%%%%%%%%
\section{RESULTS}
\label{sec:results}
%%%%%%%%%%%%%%%%%%%%%%%%%%%%%%%%%%%%%
%%%% Synthetic power grid
\subsection{Synthetic power grids}
\label{ssec:synthetic_results}

\begin{figure}[t]
\includegraphics[width=0.68\linewidth]{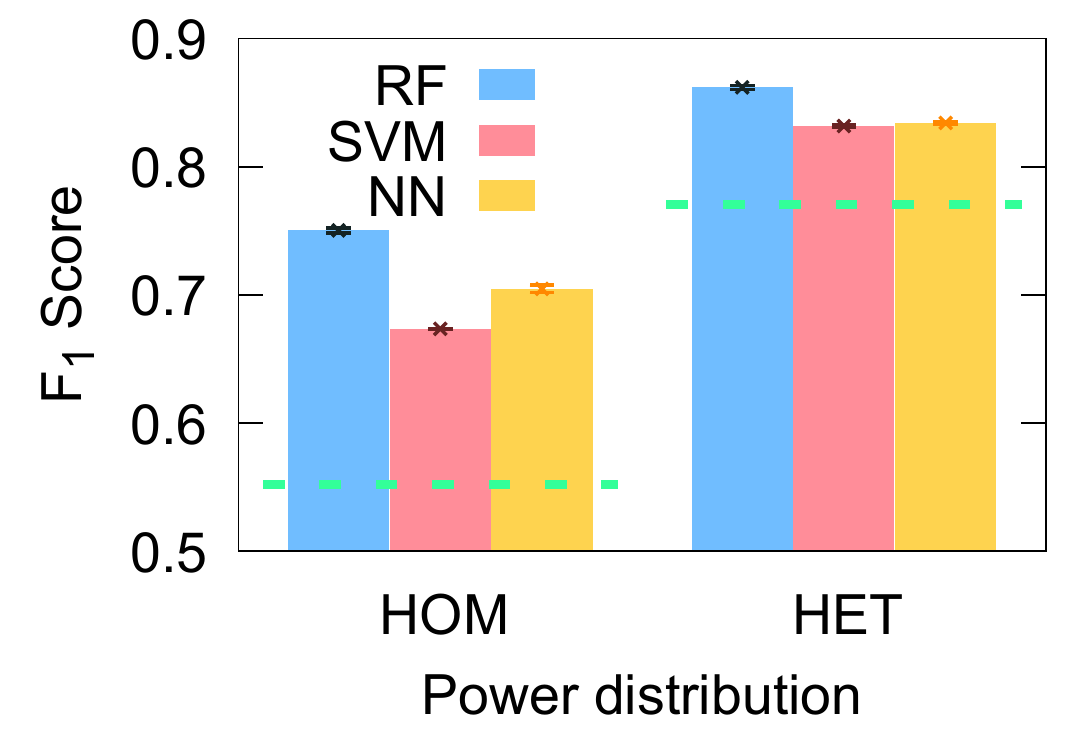}
\centering
\caption{
Average F$_1$ score for synthetic power grids with two different power distributions.
The blue bars denote the average F$_1$ score for the random forest (RF) models, pink bars for the support vector machine (SVM), and yellow bars for the artificial neural network (NN).
Three ML models trained with the \emph{heterogeneous} (HET) distribution show higher F$_1$ score values than those trained with the \emph{homogeneous} (HOM) distribution.
Among the three ML models, the RF model shows better performance than the others; thus, the RF model trained with the HET distribution is the best among the three approaches in the synthetic power grids.
The average F$_1$ score for all cases has a higher value than the Gaussian Naive Bayes prediction score $0.55$ for HOM and $0.77$ for HET, which are denoted by the green dotted lines.
The F$_1$ score is averaged over $10$ randomly sampled data sets.
}
\label{fig:f1}
\end{figure}

We evaluate the performance of our ML models with the F$_1$ score which is the harmonic mean of sensitivity and precision.
The sensitivity and precision are the predictive performance measures of ML. 
The sensitivity (also known as ``recall'') is defined as the ratio of the true positive and positively labeled data points: the precision, true positive, and positive predictions.
We present the F$_1$ score of our ML models for the HOM and HET power distributions on synthetic power grids in Fig.~\ref{fig:f1}.
All measures are averaged over ten different data sets.
In Fig.~\ref{fig:f1}, the F$_1$ score is higher than the green dotted line for both power distributions and three different ML models, where the green dotted lines denote the performance of Gaussian Naive Bayes (GNB).
The GNB is based on the Bayesian statistics~\cite{ML_book_Goodfellow} in which the likelihood of each input feature is assumed to be Gaussian and independent of each other.
We present the GNB results as the benchmark in this study.
Prediction in the HET distribution performs better than that in the HOM, and the RF shows the best performance among the three ML models.

The other performance measures---accuracy, sensitivity, precision, specificity, and negative predictive value---are summarized in Table~\ref{table:synthetic_measures}.
The accuracy is defined by the ratio with the number of truly predicted data points and the size of data sets.
The specificity (negative predictive value) corresponds to the sensitivity (precision) for the prediction of negatively labeled data.
Sensitivity and precision are the performance measures for the prediction of system recovery, and the specificity and the negative predictive value for the system failure.
Again, the trained models with the HET distribution show better performance than those with the HOM distribution.
The accuracy in Table~\ref{table:synthetic_measures} also shows that the ML models work better than the GNB prediction for both power distributions.

\begin{table}[t]
\begin{ruledtabular}
\begin{tabular}{ccccc}
HOM&RF&SVM&NN&Benchmark\\ 
\hline
Accuracy&0.756(2)&0.682(1)&0.714(1)&0.646(1)\\
Sensitivity&0.732(2)&0.654(1)&0.681(9)&0.434(3)\\
Precision&0.779(3)&0.693(1)&0.731(5)&0.755(1)\\
Specificity&0.781(4)&0.710(1)&0.748(9)&0.858(2)\\
Negative predictive value&0.744(1)&0.672(1)&0.702(3)&0.602(8)
\end{tabular}

\vspace{2ex}
\begin{tabular}{ccccc}
HET&RF&SVM&NN&Benchmark\\ 
\hline
Accuracy&0.856(3)&0.827(1)&0.829(1)&0.756(1)\\
Sensitivity&0.893(5)&0.848(2)&0.858(9)&0.819(1)\\
Precision&0.832(7)&0.813(1)&0.813(6)&0.728(1)\\
Specificity&0.81(1)&0.806(1)&0.80(1)&0.694(1)\\
Negative predictive value&0.886(4)&0.841(2)&0.851(6)&0.794(1)
\end{tabular}
\caption{
Other machine learning performance measures for synthetic power grid.
The accuracy, sensitivity, precision, specificity, and negative predictive value for machine learning prediction on the synthetic power grids.
All the measures are averaged over ten differently sampled data sets.
All measures show better performance for the \emph{heterogeneous} (HET) distribution than for the \emph{homogeneous} (HOM) distribution.
The Gaussian Naive Bayes results are presented as the performance benchmark.
}
\label{table:synthetic_measures}
\end{ruledtabular}
\end{table}

\begin{figure}[t]
\includegraphics[width=0.75\linewidth]{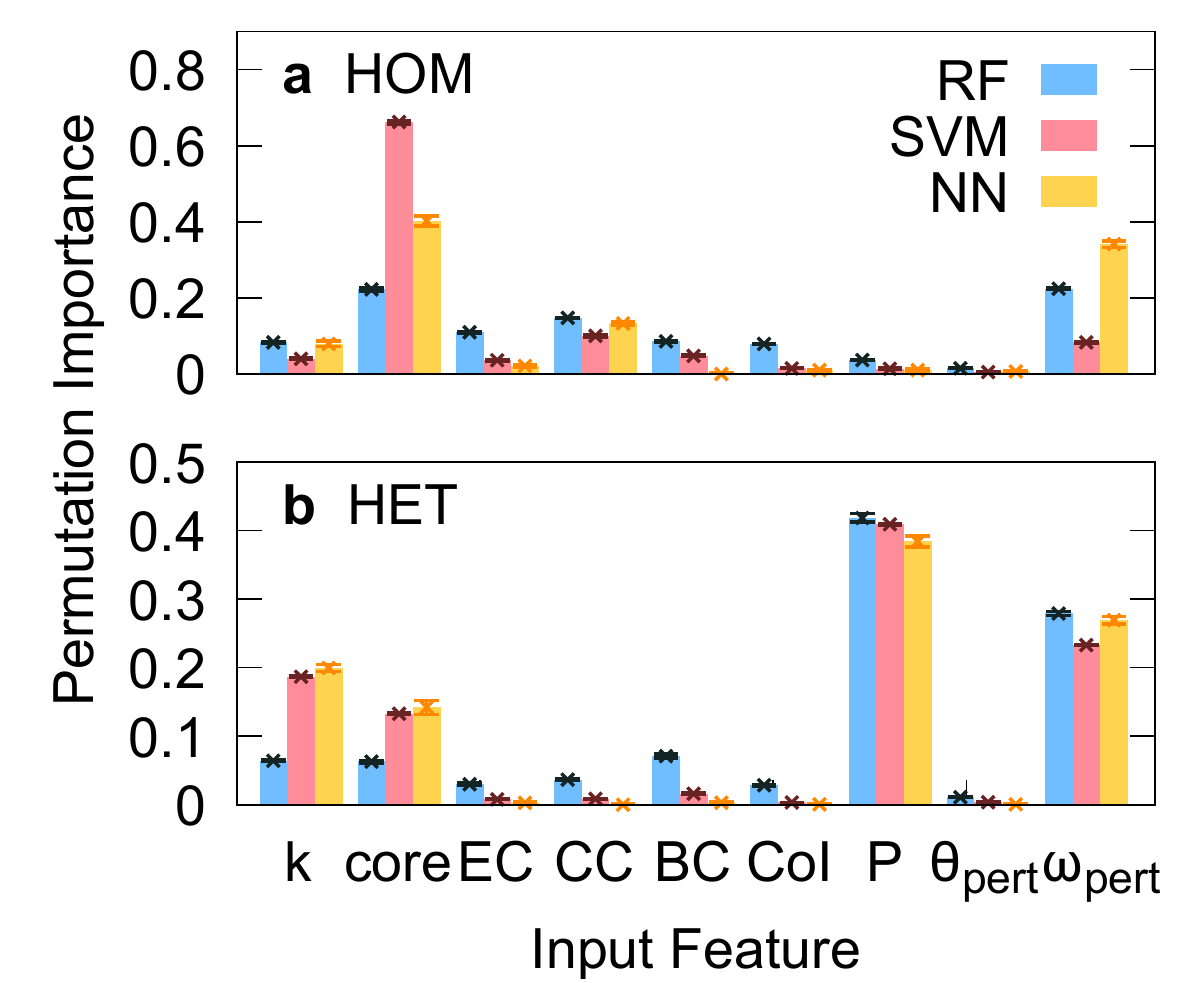}
\centering
\caption{
Permutation importance on the machine predictions of synchronization recovery from random perturbations.
The feature importance for machine learning models trained with (a) homogeneous (HOM) and (b) heterogeneous (HET) power distributions.
The blue bars denote the average permutation importance for random forest (RF) models, pink bars for support vector machine (SVM), and yellow bars for artificial neural network (NN).
The permutation importance for all input features is averaged over ten different sampled data sets.
}
\label{fig:PI}
\end{figure}

We next investigate the feature importance between nine input features measuring the permutation importance to glimpse how the ML models utilize the features.
The permutation importance enables us to estimate the dependency on each feature in the machine prediction~\cite{PIMP}.
The permutation importance for an input feature is the decrease of the accuracy when the value of the feature is randomly shuffled.
The random shuffling of the feature value breaks the correlation between the feature and the label, but the distribution of the feature is kept unchanged.
The permutation importance for all features is normalized to make the sum to unity to estimate the relative importance.
Figure~\ref{fig:PI} shows the permutation importance of nine input features for two different power distributions on the synthetic power grid.
The result shows that the ML models depend on both the dynamical and topological features.
For the ML models trained on the HOM distribution, they depend highly on $\text{core}$ and $\omega_\text{pert}$ more than other features.
SVM shows a much stronger dependency on $\text{core}$ than other predictions (Fig.~\ref{fig:PI}{\bf a}).
The ML models trained with the HET distribution also show high dependency on $\omega_\text{pert}$.
However, they show a much stronger dependency on the net power production ${P_i}$ (Fig.~\ref{fig:PI}{\bf b}).
We believe that the perturbation applied on a node with large $P$ has a larger effect on the whole system than on a node with small $P$ and the result is reflected on the machine predictions.

%%%%%%%%%%%%%%%%%%%%%%%%%%%%%%%%%%%%%
%%%% Real Power Grid
\subsection{Real power grids}
\label{ssec:real_results}

On the synthetic power grids, we train the ML models, and predict the synchronization stability estimating the prediction performance of the models.
However, applying the ML models, which are trained on synthetic power grids, to predict the stability of real power grids is another problem.
Even though the ML models work well on the training data sets, it is not guaranteed that the ML models also perform well for the different problems, in this case, different power grids of the real world.
This is the machine `transferability' issue---whether an ML model is transferable to different yet similar problems.
In this section, we test and verify the transferability of our ML models, applying them to the stability prediction of real power grids.

%%%%%%%%%%%%%%%%%%%%%%%%%%%%%%%%%%%%%
%%%% 3European Power Grid
\subsubsection{Three European power grids}
\label{sssec:ThreeEU_results}

\begin{table*}[t]
\centering
\begin{tabular}{ccccccc|ccccc}
\hline
\hline
\multirow{2}{*}{{\bf Great Britain}}&&\multicolumn{4}{c}{HOM}&&&\multicolumn{4}{c}{HET}\\ 
&&RF&SVM&NN&Benchmark&&&RF&SVM&NN&Benchmark\\
\hline
Accuracy&&0.559(4)&0.544(4)&0.573(3)&0.524(1)&&&0.865(2)&0.854(2)&0.904(3)&0.754(3)\\
Sensitivity&&0.809(6)&0.730(6)&0.788(1)&0.481(1)&&&0.931(3)&0.892(3)&0.988(4)&0.790(4)\\
Precision&&0.549(3)&0.543(3)&0.565(2)&0.544(1)&&&0.907(1)&0.928(2)&0.912(1)&0.900(1)\\
Specificity&&0.292(4)&0.344(5)&0.35(1)&0.569(1)&&&0.549(5)&0.673(7)&0.548(5)&0.5849(1)\\
Negative predictive value&&0.589(9)&0.545(6)&0.610(6)&0.506(1)&&&0.630(8)&0.569(7)&0.92(2)&0.371(4)\\
\hline
\hline
\end{tabular}

\vspace{1.5ex}
\begin{tabular}{ccccccc|ccccc}
\hline
\hline
\multirow{2}{*}{{\bf Spain}}&&\multicolumn{4}{c}{HOM}&&&\multicolumn{4}{c}{HET}\\ 
&&RF&SVM&NN&Benchmark&&&RF&SVM&NN&Benchmark\\
\hline
Accuracy&&0.633(4)&0.523(2)&0.854(6)&0.462(2)&&&0.840(3)&0.825(2)&0.876(4)&0.756(1)\\
Sensitivity&&0.602(5)&0.464(3)&0.930(8)&0.394(3)&&&0.882(4)&0.869(2)&0.972(7)&0.789(3)\\
Precision&&0.964(3)&0.977(1)&0.911(1)&0.975(1)&&&0.917(1)&0.912(6)&0.894(2)&0.898(1)\\
Specificity&&0.847(1)&0.926(3)&0.37(1)&0.929(1)&&&0.659(5)&0.639(4)&0.50(1)&0.613(6)\\
Negative predictive value&&0.236(3)&0.200(1)&0.45(2)&0.1819(5)&&&0.565(7)&0.530(3)&0.82(3)&0.492(2)\\
\hline
\hline
\end{tabular}

\vspace{1.5ex}
\begin{tabular}{ccccccc|ccccc}
\hline
\hline
\multirow{2}{*}{{\bf France}}&&\multicolumn{4}{c}{HOM}&&&\multicolumn{4}{c}{HET}\\ 
&&RF&SVM&NN&Benchmark&&&RF&SVM&NN&Benchmark\\
\hline
Accuracy&&0.735(3)&0.700(4)&0.757(1)&0.553(1)&&&0.867(4)&0.845(2)&0.882(6)&0.783(1)\\
Sensitivity&&0.783(4)&0.671(4)&0.816(1)&0.447(2)&&&0.880(5)&0.845(3)&0.970(2)&0.785(1)\\
Precision&&0.854(3)&0.908(1)&0.875(1)&0.922(1)&&&0.965(2)&0.976(1)&0.894(8)&0.962(1)\\
Specificity&&0.586(9)&0.790(2)&0.503(2)&0.883(1)&&&0.765(9)&0.844(3)&0.503(2)&0.766(4)\\
Negative predictive value&&0.466(5)&0.436(4)&0.390(2)&0.339(1)&&&0.461(8)&0.421(4)&0.795(9)&0.321(1)\\
\hline
\hline
\end{tabular}
\caption{
Other machine learning prediction performance measures for three European power grids.
Five different performance measures averaged over ten machine learning models trained on different test sets for three algorithms on three European power grids.
All measures show better performance for the \emph{heterogeneous} (HET) distribution than for the \emph{homogeneous} (HOM) distribution.
Furthermore, the prediction performance for positively labeled data is much better than that for negatively labeled data.
The benchmark denotes the results obtained from Gaussian Naive Bayes prediction.
}
\label{table:ThreeEU_measures}
\end{table*}

\begin{figure}[t]
\includegraphics[width=0.587\linewidth]{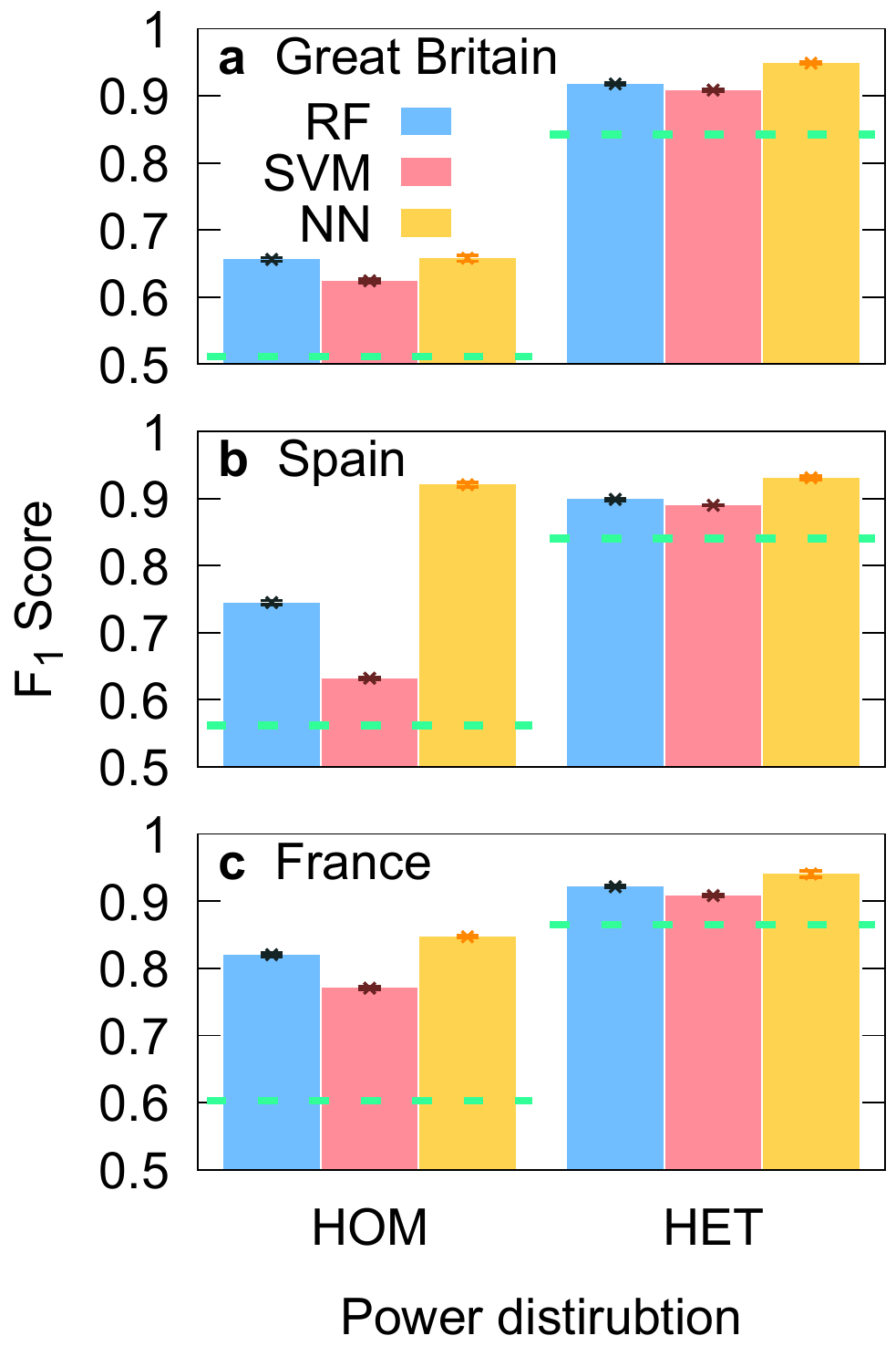}
\centering
\caption{
Average F$_1$ score for the machine prediction on three European power grids.
Machine learning performance estimated as the F$_1$ score for power grids of three European countries including (a) Great Britain, (b) Spain, and (c) France.
Like in the synthetic power grids, the machine predictions for \emph{heterogeneous} (HET) power distribution show better performance than for \emph{homogeneous} (HOM) distribution.
The green dotted lines denote the Gaussian Naive Bayes (GNB) prediction results.
The F$_1$ score of GNB results are (GB,ES,FR) = ($0.51,0.56,0.60$) for HOM and (GB,ES,FR) = ($0.84,0.84,0.86$) for HET.
All three MLs perform better than the GNB.
}
\label{fig:ThreeEU_F1}
\end{figure}

We simulate the British, Spanish, and French power grids with two power distributions, which are used in the synthetic power-grid generation and build the test data sets for three European power grids (details in Sec.~\ref{ssec:real_data}).
The ML models trained on the synthetic power grids and each power distribution are applied to predict the results from random perturbations on the real power grids, which are generated with the same power distribution.
Figure~\ref{fig:ThreeEU_F1} shows the averaged F$_1$ score of the ML models on three European power grids for each power distribution.
We also measure the other performance measures for three European power grids and the results are presented in Table~\ref{table:ThreeEU_measures}.
As in the synthetic power grids, machine predictions for the HET distribution show better performance than those for the HOM distribution.
However, the performance measures for negatively labeled data tend to have a small value than estimations for positively labeled data.
We check the prediction performance for negative data balancing data sets; however, the results show that the prediction for negative data still low.
It means that the ML models predict the recovery of the synchronous state better than the failure of power systems.

%%%%%%%%%%%%%%%%%%%%%%%%%%%%%%%%%%%%%
%%%% German Power Grid
\subsubsection{German power grids}
\label{sssec:German_result}

\begin{figure}[t]
\includegraphics[width=0.9\linewidth]{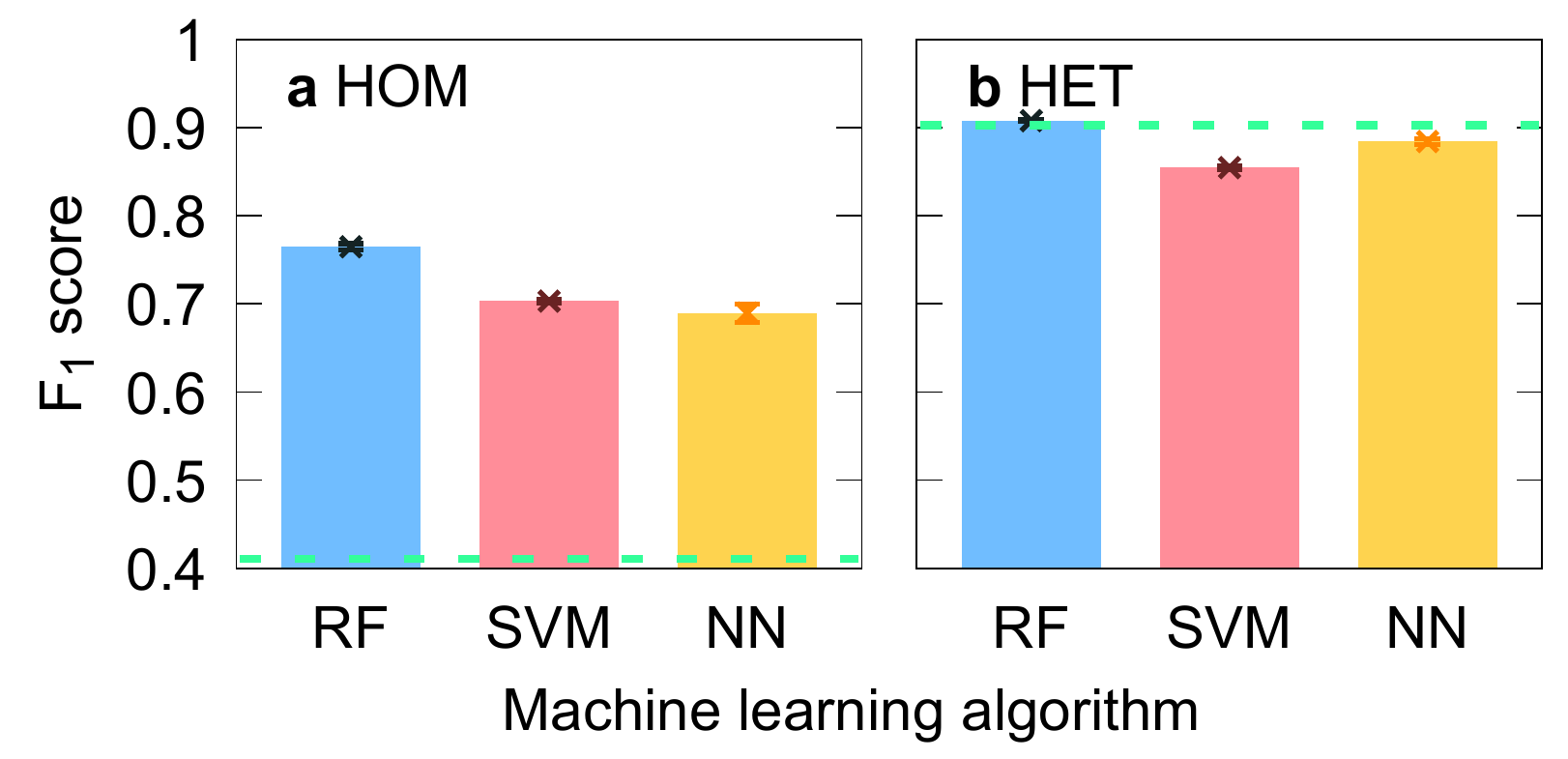}
\centering
\caption{
F$_1$ score of three different machine learning (ML) models trained with \emph{homogeneous} (HOM) and \emph{heterogeneous} (HET) power distribution for German power grid.
Three different machine learning algorithms---random forest (RF), support vector machine (SVM), and artificial neural network (NN)---trained on synthetic power grids show high prediction performance.
Three ML models trained with (b) HET distribution show better prediction performance than those trained with (a) HOM distribution.
The green dotted lines denote the score of the Gaussian Naive Bayes (GNB).
For HOM distribution, all three ML models outperform the GNB (0.41).
However, RF only shows similar performance to the GNB (0.90), but others slightly underperform than the GNB for HET distribution.
}
\label{fig:German_F1}
\end{figure}

\begin{table}[t]
\begin{ruledtabular}
\begin{tabular}{ccccc}
{\bf HOM} &RF&SVM&NN&Benchmark\\
\hline
Accuracy&0.625(4)&0.558(3)&0.55(1)&0.309(1)\\
Sensitivity&0.685(5)&0.604(4)&0.58(1)&0.277(2)\\
Precision&0.854(1)&0.842(2)&0.856(4)&0.793(1)\\
Specificity&0.227(6)&0.252(5)&0.36(2)&0.523(3)\\
Negative predictive value&0.098(2)&0.088(2)&0.114(5)&0.0984(3)
\end{tabular}

\vspace{2ex}
\begin{tabular}{ccccc}
{\bf HET} &RF&SVM&NN&Benchmark\\ 
\hline
Accuracy&0.841(2)&0.761(3)&0.802(5)&0.838(1)\\
Sensitivity&0.904(3)&0.800(4)&0.869(7)&0.869(1)\\
Precision&0.912(1)&0.914(1)&0.901(9)&0.940(1)\\
Specificity&0.423(7)&0.502(4)&0.37(1)&0.421(3)\\
Negative predictive value&0.402(5)&0.276(3)&0.298(6)&0.630(3)
\end{tabular}
\caption{
Other machine learning prediction performance measures for German power grid.
The averaged accuracy, sensitivity, precision, specificity, and negative predictive value for machine learning prediction on the synthetic power grids of two different power distributions.
HOM and HET denote \emph{homogeneous} and \emph{heterogeneous} power distributions, respectively.
The specificity (negative predictive value) is the negative prediction version of sensitivity (precision).
The prediction results of Gaussian Naive Bayes are presented as the benchmark.
}
\label{table:German_measures}
\end{ruledtabular}
\end{table}

Since there exists open power production data for the German power grid~\cite{German1}, we utilize them and build the test data set for the German power grid (details in Sec.~\ref{ssec:real_data}).
After the test data set for the German power grid is prepared, we predict whether the system recovers the synchronous state from the given perturbations with the ML models trained on the synthetic power grid.
The ML models trained with two different power distributions are applied to predict the results for the same power grid.
Figure~\ref{fig:German_F1} shows the F$_1$ score of the models and all ML models show better prediction performance than the GNB prediction for HOM distribution even when the power distribution of the German power grid is different from the training data set of the synthetic power grid.
For the HET distribution, RF shows similar performance, but the others underperform than the GNB.
Compared with the F$_1$ score for the HOM distribution, the prediction for the HET distribution shows superior performance.
The other performance measures for the HET distribution also have higher values than those for the HOM distribution (see Table~\ref{table:German_measures}).
Furthermore, the performance measures for positively labeled data have a much higher value than those for negatively labeled data similarly to our observations for the three European power grids.

We check the prediction performance of all the three ML models for the German grid, changing the ratio of positively labeled data points to negatively labeled data points in the training data sets.
When the negatively labeled data increase, specificity of the ML models also increases significantly, and the precision increases very slightly.
However, the overall prediction performance measured by accuracy decreases for all ML models.
It means that one can control the label bias in the training data set and construct the ML models to sensitively predict the negative case better than the ML models trained with balanced data set with the cost of the overall prediction performance.

\begin{figure}[t]
\includegraphics[width=0.6\linewidth]{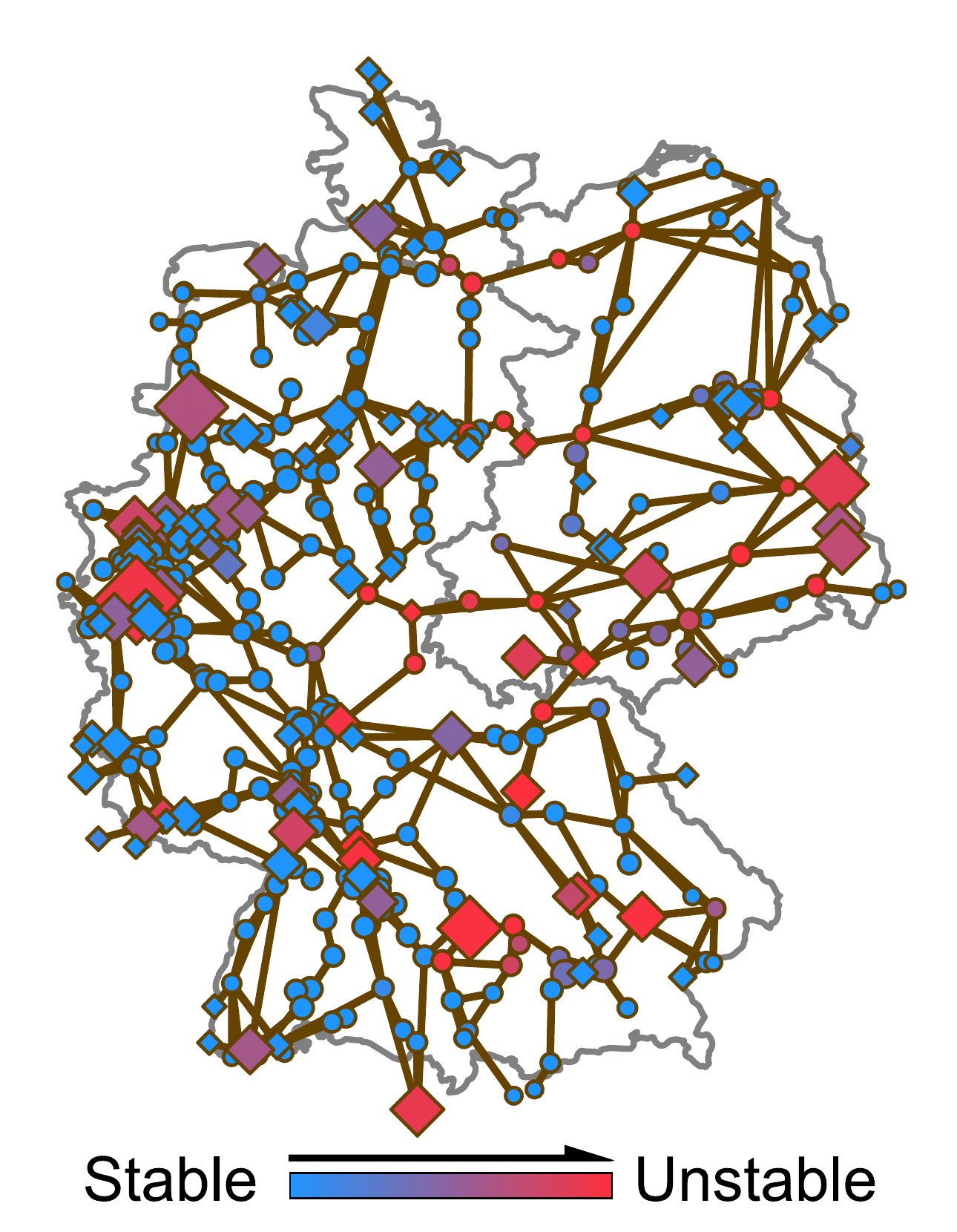}
\centering
\caption{
Basin stability of each node in German power grid.
The rhombi (circles) denote the net producers (consumers).
The size of each node is proportional to its net power production and the color of the node indicates the single-node basin stability estimated from $50$ random perturbations.
Nodes that have large net power production tend to be more unstable than others of small net power production.
Furthermore, nodes that have transmission line across the inner German border tend to be more unstable than the other nodes with similar net power production values.
}
\label{fig:German_SingleNode_BS}
\end{figure}

Based on the predictability of our ML model for the recovery of synchronous state from nodal perturbations,
we extend the prediction of the stability of the whole power grid, which can be estimated from the average basin stability.
The average basin stability is the average of single-node basin stability~\cite{BS_introduced1,BS_introduced2} for all nodes in the German power grid, and we measure the machine prediction of average basin stability as the rate of positively predicted data.
Figure~\ref{fig:German_SingleNode_BS} shows the single-node basin stability for the German power grid.
The rhombi (circles) denote the producer (consumer) nodes, and the node size represents the net power production of the node.
We color each node with its single-node basin stability value, and the red nodes are more unstable than the blue nodes.
The node with larger net power production tends to be more unstable than those that have smaller net power production.
In addition, the nodes along the German border are more unstable than the others even if they have a small amount of net power production.
In Table~\ref{table:German_BS}, we present the average basin stability of German power grids and the predictions of our ML model.
The ML predictions are averaged over ten different models trained with ten sampling data sets.
The machine predictions of basin stability are $0.697$, $0.623$, and $0.636$ under the HOM power distribution and $0.862$, $0.761$, and $0.851$ under the HET distribution for RF, SVM, and NN models, respectively.
The average basin stability also shows that the RF prediction under the HET distribution performs the best.

\begin{table}[t]
\begin{ruledtabular}
\begin{tabular}{cccc}
&RF&SVM&NN\\
\hline
HOM&0.697(5)&0.623(3)&0.59(1)\\
HET&0.862(2)&0.761(3)&0.838(7)\\ 
\hline
\ Average basin stability &\multicolumn{3}{c}{0.869} 
\end{tabular}
\caption{
Machine prediction of average basin stability of German power grid.
The average basin stability of the German power grid is estimated by averaging all nodes' single-node basin stability and it is $0.869$.
The machine prediction of basin stability is the rate of data which is predicted to be recover synchronization from given perturbation.
HOM denotes the \emph{homogeneous} and HET denotes the \emph{heterogeneous} power distribution.
The RF prediction under the HET distribution shows the best performance.
}
\label{table:German_BS}
\end{ruledtabular}
\end{table}

We check the time complexity of trained ML models comparing the numerical simulation and all three ML models demand far less time complexity than that of the numerical integration.
The numerical simulation of the AC power dynamics demands time complexity $O( T L ) \sim O( T N ) $ for each parameter, where $T$ is the number of simulation time steps, $L$ is the number of transmission lines on a power grid, and $N$ is the size of the system.
$T$ also depends on $N$ because the larger the system is, the more time the simulation needs to reach the steady state and those parameters are the main factors of large computational costs.
In this study, Eq.~\eqref{eq:swing} is numerically integrated for at least $5\times 10^8$ time steps, and $L$ and $N$ are $O(10^2)$ for power grids we dealt with.
Thus, the time complexity of the numerical simulation is $O(10^{10})$ for each parameter.
For the ML approaches, we use in this study; however, once the training of ML model is finished, the ML model demands time complexity which is independent of $T$ and $N$.
The time complexity of three algorithms are $O(N_\text{tree}~N_\text{depth})$, $O(N_\text{input}~N_\text{vector})$, and $O(\sum_i N^{i}_\text{layer}~N^{(i+1)}_\text{layer})$ for RF, SVM, and NN, respectively, where $N_\text{tree}$ is the number of decision trees, $N_\text{depth}$ is the maximum depth of decision tree, $N_\text{input}$ is the number input features, $N_\text{vector}$ is the number of support vectors, and $N^{i}_\text{layer}$ is the size of $i$th layer.
We use $N_\text{tree}=500$ decision trees for RF model and we get the average maximum depth of decision trees $N_\text{depth, Homo} = 68.6$ for HOM and $N_\text{depth, Hetero} = 63.2$ for HET after the training is finished.
Thus, we get the complexity of RF is $O(10^4)$.
For SVM, we get the average number of support vectors $N_\text{vector, Homo}=189~442.7$ for HOM and $N_\text{vector, Hetero}= 86~507.5$ for HET.
Finally, we can calculate the complexity of our SVM model as $O(10^6)$.
The structure of NN model is $11\times 21\times 21 \times 11 \times 2$ and $N_\text{input}+1=9+1$ features (nine input features and one bias) are fed to the NN model.
Thus, we get the time complexity $O(10^4)$ for NN.

%%%%%%%%%%%%%%%%%%%%%%%%%%%%%%%%%%%%%
%\section{Summary and Discussions}
\section{CONCLUSTION AND DISCUSSIONS}
\label{sec:conc_disc}

In this work, we have investigated the applicability of ML techniques to predict the power-grid stability developing transferable ML models.
The transferable ML models (RF, SVM, and NN) are trained on synthetic power grids using both topological and dynamical features.
The synthetic power grids were generated with the random growth model~\cite{netgen}, and we assigned two different power distributions, the HOM and HET distributions, to the training data sets.
The performances of our ML models have been estimated with average accuracy, sensitivity, precision, specificity, negative predictive value, and mainly the F$_1$ score.
All ML models have shown high predictive performance for both power distributions achieving the F$_1$ score higher than the score of GNB.
In particular, the RF models trained with the HET distribution have shown the best performance of F$_1$ score higher than $0.85$ for the synthetic power grids.
We have tried to understand how our ML models work by measuring the permutation importance and we have found that both topological and dynamical features play an important role in machine prediction.
The $\text{core}$ is the most important feature in the predictions for the HOM power distribution, and the net power production plays the most important role in the prediction performance for the HET distribution.
We believe that the topological features are more important than the net power for HOM distribution, but the net power is much more important to predict the recovery of synchronous state for HET distribution than other features.
Fortunately, all the ML models give the predictions in an understandable way.

We have also estimated the prediction performance of transferable ML models on real power grids with the performance measures and especially with the F$_1$ score.
We have simulated four European power grids, including Great Britain, France, Spain, and Germany, and constructed the test data sets for the real power grids.
Surprisingly, even though our models are trained on synthetic power grids, the models have shown high prediction performance on real power grids.
Comparing two power distributions, the ML models trained with the HET distribution performed better than those with the HOM distribution.
The HOM distribution is quite simple to capture the effect of the real net power distribution on the system stability.
The HET distribution is more complex than the HOM distribution, and the ML model trained with HET distribution could learn the data of various cases.
We believe that the ML models for HET distribution should perform better than those for HOM distribution.
In addition, all models better predict the recovery of synchronous state from random perturbations showing high value of the sensitivity and the precision than to predict the failure of recovery, which is estimated with the specificity and the negative predictive value.
For the German power grid, the RF models show the highest average F$_1$ score, but for the other three European countries, NN models show better performance with the RF models.
The NN models trained with the HOM distribution on the synthetic power grid show the best performance on the Spanish power grid.
The machine prediction of whole power-grid stability has also been discussed in this paper.
We have predicted the overall stability of the power grid estimating the average basin stability with our ML models and the RF models trained with the HET distribution give the closest value to the average basin stability obtained from the numerical simulation on the German power grid.

Even if the bias in the training data set is removed, the ML models predict better system recovery on the real power-grid topologies than the system failure.
This result is obvious in the prediction of the ML models trained with HET distribution.
We believe that this property of our ML models is beneficial to apply the ML techniques to real power system control and operation.
When a power malfunction occurs, cascading failure happens and the entire system fails in a few minutes~\cite{Fail_Time}.
Our ML models give prediction results of the malfunction in a few seconds and can alert the danger quickly.
If the machine prediction is recovery, people can have stronger trust in the result.
In the opposite case of failure prediction, however, system operators should pay more attention and need to figure out whether the malfunction ultimately leads to the system failure or recovery.
In this manner, the machine predicts the safety of the power system conservatively and can help system operators.

The ML techniques have much potential applicability.
Power-grid stability prediction can be regarded as a classification problem and the problem is handled as such in this paper.
However, the cascading failure of a power system, which is also an important problem in the system stability analysis~\cite{Turkish,vulnerability1,vulnerability3,EU,cascading,cascading_powerlaw}, belongs to a different group of problems. 
The cascading failure is a dynamic phenomenon and this problem should be dealt with other ML techniques because the state of the system changes in time.
To deal with such time varying phenomena, the Recurrent Neural Network (RNN) can be a candidate for the ML technique.
The typical example of RNN is the reservoir computing method, and the method is used to predict the trajectory of a chaotic system recently~\cite{ML_RC,ML_Forecast,ML_ComplexSystem}.

\begin{acknowledgments}
This research was supported by ``Human Resources Program in Energy Technology" of the Korea Institute of Energy Technology Evaluation and Planning (KETEP), granted financial resource from the Ministry of Trade, Industry and Energy, Republic of Korea Grant No. 20194010000290 (S.-G.Y. and B.J.K.),
the National Research Foundation of Korea through the Grant No. NRF-2020R1A2C2010875 (S.-W.S.),
the National Agency of Investigation and Development, ANID, of Chile through the grant FONDECYT No. 11190096 (H.K.), the Korea Institute of Energy Technology (KENTECH) No. KRG2021-01-003 (H.K.).
This work was also partly supported by Institute of Information \& Communication Technology Planning \& Evaluation grant funded by the Korea government (No. 2020-0-01343, Artificial Intelligence Convergence Research Center (Hanyang University)) (S.-W.S.).
S.-G.Y. was also supported by an appointment to the YST Program at the APCTP through the Science and Technology Promotion Fund and Lottery Fund of the Korean Government. This was also supported by the Korean Local Governments - Gyeongsangbuk-do Province and Pohang City.
\end{acknowledgments}

\section*{AUTHOR DECLARATIONS}
\subsection*{Conflict of Interest}
The authors have no conflicts of interest to declare.

\section*{DATA AVAILABILITY}
The data that support the findings of this study are available from the corresponding author upon reasonable request.
All data sets and Python codes of the three machine learning algorithms described in this research may be found in GitHub at \url{https://github.com/totisviribus/ML_PowerGrid.git}, Ref.~\onlinecite{my_git}.

\section*{REFERENCES}

\iffalse
\bibliography{reference}
\fi

%aipnum4-2.bst 2019-01-14 (MD) hand-edited version of apsrev4-1.bst
%Control: key (0)
%Control: author (8) initials jnrlst
%Control: editor formatted (1) identically to author
%Control: production of article title (0) allowed
%Control: page (1) range
%Control: year (1) truncated
%Control: production of eprint (0) enabled
%

\end{document}